\documentclass[pra, twocolumn]{revtex4}

\usepackage{latexsym}
\usepackage{amssymb, amsmath}
\usepackage{exscale}
\usepackage{bm, graphics}
\usepackage{graphicx, subfigure, pstricks}

\newcommand{\D}{\mbox{\rm d}}
\newcommand{\uhb}[1]{\underline{\hat{\mathbf{#1}}}}
\newcommand{\hb}[1]{\hat{\mathbf{#1}}}
\newcommand{\uh}[1]{\underline{\hat{#1}}}
\newcommand{\ket}[1]{$\left|#1\right\rangle$}

\usepackage{tikz}
\usetikzlibrary{arrows}

\begin{document}
\title{Cavity-assisted spontaneous emission
of a single
$\Lambda$-type
emitter
as a source of single-photon packets with controlled shape
}

\author{M. Khanbekyan}
\email[E-mail address: ]{mikayel.khanbekyan@ovgu.de}
\affiliation{Institut f{\"u}r Theoretische Physik, Universit{\"a}t
  Magdeburg, Postfach 4120, D-39016 Magdeburg, Germany}

\author{D.-G. Welsch}
\affiliation{Theoretisch-Physikalisches Institut,
Friedrich-Schiller-Universit\"at Jena, Max-Wien-Platz 1,
D-07743 Jena, Germany}

\date{\today}

\begin{abstract}
The radiation emitted by
a classically pumped
three-level $\Lambda$-type
emitter in a resonator cavity featuring
both radiative and unwanted losses
is studied.
In particular, the
efficiency of one-photon Fock state excitation of the
outgoing wave packet
and
the
spatiotemporal shape
of the wave packet
are investigated.
It is shown that for given
shape
of the emitter--cavity interaction,
adjusting the
shape
of the pump pulse 
renders it possible
to
generate the emission of one-photon Fock
state
wave packets of
desired shape with high efficiency.

\end{abstract}


\maketitle

\section{Introduction}
\label{introduction}
The interaction of a single emitter with
the
quantized radiation-field
in a high-$Q$ cavity
serves as a basic ingredient in various
schemes in quantum information
science (for a review see, e.\,g.,
Ref.~\cite{reiserer:1379}).
In this context, an essential prerequisite
for quantum information processing~\cite{knill:46} is the deterministic
generation of
indistinguishable
light pulses of one-photon Fock states
on demand with well-controlled features, such as efficiency,
frequency, polarization, timing etc..
Among the various features of single-photon emission, control
and manipulation of the spatiotemporal shape of the single-photon
light pulses is essential for quantum network applications where
symmetric wavepackets are required for a time reversal of the emission
process~\cite{cirac:3221} or where time-bin entanglement is
used~\cite{marcikic:062308}.

The most prominent way to generate single-photon Fock states in
high-finesse optical cavities interacting with
single emitters is based on vacuum stimulated Raman adiabatic approach
(vSTIRAP) technique~\cite{law:2067, bergmann:1003}.
A single transition of
an emitter with a $\Lambda$-type energy level scheme is excited by
a pump laser stimulating the emission into a
strongly coupled cavity--emitter transition. Alongside the
principal
realizations of vSTIRAP in atomic systems (see
e.g.~\cite{specht:469}), a $\Lambda$-type scheme can be also implemented in
semiconductor quantum dot--microcavity systems, where
a $\Lambda$-type system is created by applying
a strong magnetic field to a singly charged quantum dot resulting in
Zeeman splitting of the spin states in the conduction
band~\cite{kiraz:032305}
or by applying a lateral electric field which charges a single
quantum dot with a single electron resulting in optical transitions
between ground and excited electron states and the charged
exciton~\cite{jaritz:29}.
In particular, in atomic systems control of
the photon frequency~\cite{muecke:063805},
wavepacket
length~\cite{muecke:063805} and phase~\cite{specht:469} have been
reported.
The efficiency of the generation of single-photon Fock states in
a scheme based on vSTIRAP technique
is governed by an interplay of various system parameters,
such as the transition dipole matrix element, intensity and
amplitude of the pump pulse, frequency detuning etc.
Further, it has also been reported that the spatiotemporal shapes of
the outgoing wave packets of the emitted
one-photon Fock state radiation
follow the
driving laser pulse shapes~\cite{kuhn:067901, keller:1075}.
However, this behavior is only observed
for a certain range of parameters
and the efficiency of
one-photon Fock state generation is very low, which makes it impossible
to use the scheme for further applications in quantum
communications.
More recently,
the possibility of generation of single-photon wave packets
with arbitrary shapes in the case when
the atom--cavity interaction remains constant during the
generation process has been demonstrated
both theoretically, on using
the master equation formalism for describing
the atom--photon system~\cite{vasilev:063024}, and
experimentally, with un-trapped atoms injected into a cavity by means of an
atomic fountain~\cite{nisbet-jones:103036}.

In an earlier work, we have studied the potential of
a single two-level atom in a high-$Q$ cavity as a
single-photon emitter, where the wave packet
associated with the emitted photon is
shorter than the cavity decay
time~\cite{khanbekyan:013822,fidio:043822}.
Within the frame of exact quantum electrodynamics,
we have particularly shown the feasibility of
the generation of single-photon wave packets with
time-sym\-metric spatiotemporal shapes.

In the present article we
extend the theory to
the interaction of a pumped three-level
$\Lambda$-type emitter with a realistic cavity-assisted
quantized electromagnetic
field, with the aim of
well-defined and controlled
one-photon Fock-state emission.
In particular, we study in detail both the properties of the
excited outgoing wave packet
and the efficiency to prepare it in a single-photon Fock state.
The
general explicit expression
derived
for the spatiotemporal shape
of the excited outgoing wave packet allows us to study different
coupling
regimes of the emitter--pump and the emitter--cavity-field interactions.
In this context,
we show the possibility to generate single-photon wave packets
with requested spatiotemporal shapes,
which
can be achieved by
using
appropriately designed
driving laser pulses.

The paper is organized as follows. The basic equations for the
resonant interaction of an emitter with a cavity-assisted
electromagnetic field are given in Sec.~\ref{sec2}. In Sec.~\ref{sec3}
the Wigner function of the quantum state the excited outgoing wave
packet is prepared in and the spatiotemporal shape of the wave packet are studied
for the case of a classically pumped three-level $\Lambda$-type
emitter. Section~\ref{sec7} is devoted to the problem of
generating single-photon wave packets of desired
shapes.
A summary and some concluding remarks are given in Sec.~\ref{sec9}.


\section{Basic equations}
\label{sec2}

We consider a single emitter (position ${\bf r}_A$)
that interacts with the electromagnetic field
in the presence of a dispersing and absorbing dielectric medium with a
spatially
varying and frequency-dependent complex permittivity
\begin{equation}
    \label{1.0}
      \varepsilon({\bf r},\omega) = \varepsilon'({\bf r},\omega)
      + i\varepsilon''({\bf r},\omega),
\end{equation}
with the real and imaginary parts $\varepsilon'({\bf r},\omega)$
and $\varepsilon''({\bf r},\omega)$, respectively.
Applying the multipolar-coupling scheme in electric dipole
approximation, we may write the Hamiltonian
that governs the temporal evolution
of the overall system, which consists of the electromagnetic
field, the dielectric medium (including the dissipative degrees
of freedom), and the emitter coupled to the field,
in the form of (for details, see Refs.~\cite{knoell:1, vogel})
\begin{align}
   \label{1.1}
        \hat{H} =
&
        \int\! \D^3{r} \int_0^\infty\! \D\omega
      \,\hbar\omega\,\hb {f}^{\dagger}({\bf r },\omega)\cdot
      \hb{ f}({\bf r},\omega)
\nonumber\\&
+
	  \sum _k
        \hbar
\omega _{kk}
\hat{S} _{kk}
-
   \hb{ d}_A\cdot
        \hb{E}({\bf r}_A)
.
\end{align}
In this equation, the first term is the Hamiltonian of
the \mbox{field--me}\-dium system, where the
fundamental
bosonic fields
\mbox{$\hb{ f}({\bf r},\omega)$}
and \mbox{$\hb{f}^\dagger({\bf r},\omega)$},
\begin{align}
    \label{1.3}
    &      \bigl[\hb{f} ({\bf r}, \omega),
      \hb{f} ^{\dagger } ({\bf r }',  \omega ') \bigr]
      = 
      \delta (\omega - \omega  ')
      \bm{\delta}
      ({\bf r} - {\bf r }') ,
\\
\label{1.3-1}
&\bigl[\hb{f} ({\bf r}, \omega),
      \hb{f} ({\bf r }',  \omega ') \bigr]
= \bm{0},
\end{align}
play the role of the canonically conjugate system variables.
The second term is the Hamiltonian of the emitter, where
the $\hat{S}_{kk'}$ are the flip operators,
\begin{equation}
   \label{1.5}
   \hat{S} _{kk'} =
   | k\rangle
   \!\langle k' |
,
\end{equation}
corresponding to the \mbox{\ket{k} $\!\leftrightarrow$ \!\ket{k'}} transition
with the frequency $\omega_{kk'}$,
where
$|k\rangle$
is the energy eigenstates
of the emitter. Finally, the last term is the emitter--field
coupling energy, where
\begin{equation}
   \label{1.7}
    \hb{ d}_A = \sum _{kk'}
    {\bf d} _{Akk'}  \hat{S} _{kk'}
\end{equation}
is the electric dipole-moment operator
($ {\bf d} _{Akk'}$ $\!=$ $\!\langle k|
\hb{ d}_{\!A} | k' \rangle$),
and the operator of the
medium-assisted electric field $\hb{E}({\bf r})$
can be expressed in terms of the variables
$\hat{\mathbf{f}}(\mathbf{r},\omega)$ and
$\hat{\mathbf{f}}^\dagger(\mathbf{r},\omega)$ as
follows:
\begin{equation}
\label{1.9}
\hb{E}({\bf r}) = \hb{E}^{(+)}({\bf r})
        +\hb{E}^{(-)}({\bf r}),
\end{equation}
\begin{equation}
\label{1.10}
\hb{E}^{(+)}({\bf r}) = \int_0^\infty \D\omega\,
      \uhb{E}({\bf r},\omega),
\quad
\hb{E}^{(-)}({\bf r}) =
[\hb{E}^{(+)}({\bf r})]^\dagger,
\end{equation}
\begin{multline}
      \label{1.11}
      \uhb{ E}({\bf r},\omega) =
\\
i \sqrt{\frac {\hbar}{\varepsilon_0\pi}}\,
\frac{ \omega^2}{c^2}
      \int \D^3r'\sqrt{\varepsilon''({\bf r}',\omega)}\,
      \mathsf{G}({\bf r},{\bf r}',\omega)
      \cdot\hb{f}({\bf r}',\omega).
\end{multline}
In the above, the classical (retarded)
Green tensor $\mathsf{G}({\bf r},{\bf r}',\omega)$
is the solution to the equation
\begin{equation}
      \label{1.13}
      \bm{\nabla}
\times
    \bm{\nabla}\!  \times \mathsf{G}  ({\bf r }, {\bf r }', \omega)
      - \frac {\omega ^2 } {c^2} \,\varepsilon ( {\bf r } ,\omega)
      \mathsf{G}  ({\bf r}, {\bf r }', \omega)
      =  \bm{\delta} ^{(3)}  ({\bf r }-{\bf r }')
      \end{equation}
together with the boundary condition at infinity,
$\mathsf{G}({\bf r},{\bf r}',\omega)\to 0$ if
$|\mathbf{r}-\mathbf{r}'|\to\infty$, and defines the structure of the
electromagnetic field
formed by
the present dielectric bodies.

\section{
Single-photon generation efficiency and wave packet shape}
\label{sec3}
\subsection{
Pumped three-level
$\Lambda$-type emitter in a cavity}
\label{sec3.3}

\begin{figure}[t!]
        \includegraphics[width=0.48\textwidth]{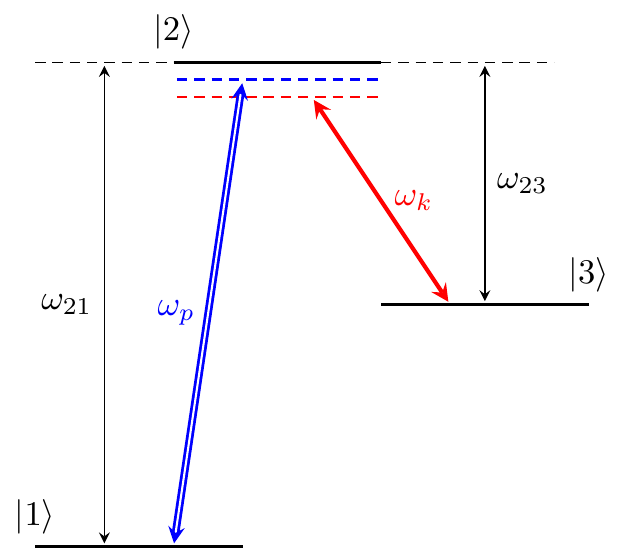}
        \caption{
          \label{fig1}
          Scheme of energy levels and transitions for the three-level
          $\Lambda$-type atomic emitter.}
      \end{figure}

Let us consider a single atom-like emitter placed in a resonator cavity
and assume that only a single transition
(\mbox{\ket{2} $\!\leftrightarrow$ \!\ket{3}}, frequency $\omega_{23}$)
is quasi resonantly
coupled to a narrow-band cavity-assisted electromagnetic field
(frequency $\omega_k$), cf. Fig.~\ref{fig1}.
We further assume, that an external (classical) pump field with
quasi resonant frequency $\omega_p$ and
(time-dependent)
Rabi frequency
$\Omega_p(t)$ is applied to the
\mbox{\ket{1} $\!\leftrightarrow$ \!\ket{2}}
transition (frequency $\omega_{21}$), cf. Fig.~\ref{fig1}.
For the sake of simplicity of presentation, we restrict our treatment to
the one-dimensional case ($z$ axis) and assume that the resonator cavity
is formed by an empty body
bounded
by an outcoupling fractionally transparent mirror at $z=0$ and
a perfectly reflecting mirror at $z=-l$,
and (negative) $z_A$ is the position of the emitter
inside the cavity.

In this case,
the  one-dimensional version of the Hamiltonian~(\ref{1.1}) in the
rotating-wave approximation reads
\begin{align}
   \label{1.15}
        \hat{H} =
&
        \int\! \D z\int_0^\infty\! \D\omega
      \hbar\omega\hat {f}^{\dagger}(z, \omega)
      \hat{f}(z, \omega)
+ \hbar \omega _{21} \hat {S}_{22} + \hbar \omega _{31} \hat {S}_{33}
\nonumber\\&
  -g(t)\left[
        d_{23}\hat{S}_{32}^{\dagger}
            \hat{E}^{(+)}(z_A)
            +
\mbox{H.c.} \right]
\nonumber\\[.5ex]
&
-
\frac{\hbar}{2}
\Omega_p
(t)\left[
        \hat{S}_{12}^{\dagger}
            e^{-i\omega_p t}
            +
\mbox{H.c.} \right],
\end{align}
with \mbox{$\omega _{31} $ $\!=$ $\omega _{21}- \omega _{23}$}.
Here,
the (real) time-dependent
function $g(t)$ defines the (time-dependent)
shape
of the
interaction of the emitter with the cavity field, which without loss
of generality can be
chosen to be
normalized to unity.
It may be used to
simulate the (quasistatic) motion of the emitter through the
cavity in the direction perpendicular to the cavity axis and/or motion
along the cavity axis.

In what follows we assume that
the emitter is initially (at time $t$ $\!=$ $\!0$)
prepared in the state \ket{1} and the rest of
the system, i.\,e., the part of the
system that consists of the
electromagnetic field and the dielectric media
(i.\,e., the cavity),
is prepared in the ground
state \ket{\{0\}}.
Since, in the case under consideration,
we may approximately span the Hilbert space of the whole system by the
single-excitation states,
we may expand the state
vector of the overall system at later times $t$ ($t\ge 0$) as
\begin{multline}
\label{1.17}
	|\psi(t)\rangle =
   C_1(t)
  |\!\left\lbrace0\right\rbrace\!\rangle|1\rangle +
  C_2(t)
e^{-i\omega_{21} t}
  |\!\left\lbrace0\right\rbrace\!\rangle |2\rangle
\\[.5ex]
+\!\int\! \D z \int_0^\infty \!\D \omega\,
   C_3(z, \omega, t)
   e^{-i(\omega +\omega _{31})t}
   \hat{f}^{\dagger}(z, \omega)
   |\!\left\lbrace0\right\rbrace\!\rangle |3\rangle ,
\end{multline}
where $\hat{f}^{\dagger}(z, \omega)|\{0\}\rangle$ is
an excited single-quantum state of the combined
field--cavity system.

It is not difficult to prove
that the Schr\"odinger equation for
\ket{\psi(t)}
leads to the following system of
differential
equations for the probability amplitudes
$C_1(t)$, $C_2(t)$ and $C_3(z, \omega, t)$:
\begin{equation}
  \label{1.19}
  \dot {C_1} =
      \frac{i}{2}
\Omega_p
(t)
      e^{i\Delta_pt}
      C_2(t),
\end{equation}
\begin{multline}
 \label{1.20}
 \dot {C_2} =
 \frac{i}{2}
\Omega_p
(t)
      e^{-i\Delta_pt}
      C_1(t)
\\[.5ex]
      -\frac{d_{23}}{\sqrt{\pi \hbar \varepsilon _0  \mathcal{A}}}
      \int_0^\infty\! \D\omega\, \frac{\omega ^2}{c^2}
      \int \D z
      \sqrt{\varepsilon''(z,\omega)}\,
\\[.5ex]
 \times
      G(z_A, z,\omega)
      C_3(z, \omega, t)g(t)
      e^{-i (\omega - \omega_{23})t},
\end{multline}
\begin{multline}
  \label{1.18}
 \dot {C_3}(z, \omega, t) =
      \frac{d_{23}^*}{\sqrt{\pi \hbar \varepsilon _0  \mathcal{A}}}
      \frac{\omega ^2}{c^2}
       \sqrt{\varepsilon''(z,\omega)}\,
\\[.5ex]
\times
      G^*(z_A, z,\omega)
      C_2(t)g(t)
      e^{i (\omega - \omega_{23})t},
\end{multline}
where $\mathcal{A}$ is the area
of the coupling mirror of the cavity,
and $\Delta_p=\omega_p-\omega_{21}$
is the detuning of
the pump frequency
from the \mbox{\ket{1} $\!\leftrightarrow$ \!\ket{2}} transition
frequency.
The Green function $G(z,z',\omega)$ determines the spectral response
of the resonator cavity. For a
sufficiently high-$Q$ cavity, the excitation spectrum effectively turns into
a quasi-discrete set of lines of mid-frequencies $\omega_k$
and widths $\Gamma_k$, according to the poles of the
Green function at the complex frequencies
\begin{equation}
  \label{1.21}
    \tilde{\omega}_k
    = \omega_{k}
      - {\textstyle\frac {1} {2}}i\Gamma _{k},
\end{equation}
where the linewidths are much smaller than the
line separations,
\begin{equation}
\label{1.30-1}
\Gamma_k \ll {\textstyle\frac{1}{2}}(\omega_{k+1} - \omega_{k-1}).
\end{equation}
In this case, assuming that the $k$th mode of the cavity is quasi
resonantly coupled to the transition
\mbox{\ket{2} $\!\leftrightarrow$ \!\ket{3}} with the transition
frequency $\omega_{23}$
(cf. Fig.~\ref{fig1}),
we substitute the formal solutions to Eqs.~(\ref{1.19}) and (\ref{1.18})
[with the initial condition \mbox{$C_1(0)=1$}, \mbox{$C_2(0)=0$} and
\mbox{$C_3(z, \omega, 0)=0$}] into  Eq.~(\ref{1.20}).
Then, using
the
relations for the Green tensor $G(z_1, z_2, \omega)$
as given by
Eqs.~(\ref{app:1}) and (\ref{app:2}),
we can
derive the integro-differential equation
\begin{align}
 \label{1.22}
 \dot {C_2} =
 \frac{i}{2}
\Omega
_p(t)
      e^{-i\Delta_pt}
 +
\int_0^t \! \D t'\,
    K(t,t')
    C_2(t'),
\end{align}
where the kernel function $K(t,t')$ reads
\begin{align}
  \label{1.24}
  K(t,t')=
&
  -\frac{1}{4}
\Omega
_p(t)
\Omega
_p(t')
    e^{-i\Delta_p(t-t')}
\nonumber\\[1ex]&
    -
    \frac{1}{4}
   \alpha_k 
   \tilde{\omega}_k
   g(t)
    g(t')
    e^{-i(\Delta_k - i \Gamma_k/2)(t-t')},
\end{align}
with \mbox{$\Delta_k$ $\!=\omega_k$ $\!-\omega_{23}$} and
\begin{equation}
  \label{1.23}
   \alpha_k  =  \frac{4|d_{23}|^2 }
    {\hbar \varepsilon _0  \mathcal{A}
l}\,
    \sin^2
    (\omega_k
    z_A/c)
.
\end{equation}
From Eq.~(\ref{1.22}) together with Eq.~(\ref{1.24})
we can conclude
that
\mbox{$R_k$ $\!\equiv\sqrt{\alpha_k\omega_k}$}
can be
regarded as vacuum Rabi frequency of emitter--cavity interaction.

Following Ref.~\cite{khanbekyan:013822}, it can be shown that
when the Hilbert space of the system is
effectively spanned by a single excitation,
on a time-scale that is short compared to the inverse spontaneous
emission rate of the emitter,
the multimode Wigner function of the quantum state
of the outgoing field can be derived to be 
(Appendix \ref{app2})
\begin{equation}
\label{1.25}
W_{\mathrm{out}} (\alpha_i, t)
= W_1(\alpha_1,t)
\prod_{i\neq 1}
     W_i^{(0)}(\alpha _i, t),
\end{equation}
where
\begin{equation}
\label{1.27}
W_1(\alpha,t)
= [1-\eta(t)]W_1^{(0)}(\alpha)
     +\eta(t)W_1^{(1)}(\alpha),
\end{equation}
with $W_i^{(0)}(\alpha)$ and $W_i^{(1)}(\alpha)$
being the Wigner functions of the vacuum
state and the one-photon Fock state, respectively, for the
$i$th nonmonochromatic mode. As we see, the
mode labeled by the subscript \mbox{$i$ $\!=$
$\!1$},
is
basically
prepared
in a mixed state of a one-photon Fock state and the
vacuum state, due to
unavoidable existence of unwanted losses.
The other nonmonochromatic modes of the
outgoing field with \mbox{$i$ $\!\neq$ $\!1$} are in the vacuum
state and, therefore, remain unexcited.

The Wigner function $W_1(\alpha,t)$
of the mode
associated with the excited outgoing wave packet
reveals that
$\eta(t)$ can be regarded as being the efficiency
to prepare the excited outgoing wave packet in
a one-photon Fock state (Appendix \ref{app2}):
\begin{equation}
 \label{1.29}
         \eta(t)
=
\int_0^{\infty}\!
         \D\omega\, |F(\omega ,t)| ^2
\simeq
\int_{-\infty}^\infty
         \D\omega\, |F(\omega ,t)| ^2
,
 \end{equation}
with
\begin{multline}
  \label{1.30}
    F(\omega, t)=
    \frac{d_{23}}{\sqrt{\pi \hbar \varepsilon _0  \mathcal{A}}}
    \sqrt{\frac{c}{\omega}}
    \frac{\omega^2}{c^2}
\\[1ex]\!\!\!
\times\!
\!\!      \int ^t _0\! \D t'
     G^*(0^+, z_A, \omega)
     C_2^*(t')g(t') e^{i\omega(t-t')}
     e^{i
\omega
_{23}t'}
 e^{i\omega_{31}t}
,\!\!\!\!
\end{multline}
where $0^+$ indicates the position \mbox{$z$ $\!=0$} outside the cavity.
The excited outgoing wave packet
(mid-frequency $\omega_k$)
is characterized by the mode function
\begin{equation}
  \label{1.31}
    F_1(\omega,t)
    = \frac{F(\omega , t)}{\sqrt{\eta(t)}}\,
.
\end{equation}
The intensity of the outgoing field at position $z$ is given by
\begin{equation}
  \label{1.33}
   I(z,t)
  =
  \eta(t)
  |\phi_1(z,t)|^2
,
\end{equation}
with
\begin{equation}
  \label{1.35}
\phi_1(z, t)
    =
    \frac{1}{2}
     \int_0^{\infty}\!
     \D\omega\,
      \sqrt{\frac{\hbar\omega}{\varepsilon_0 c \pi\mathcal{A}
        }}\,
     e^{-i \omega z/c}
    F_1(\omega, t)
.
\end{equation}
Equation~(\ref{1.33}) reveals that $\phi_1(z,t)$
represents the spatiotemporal shape of the
outgoing field associated with the excited nonmonochromatic mode
$F_1(\omega,t)$.

\subsection{Efficiency of single-photon Fock state generation}
\label{sec5}

%
Inserting Eq.~(\ref{1.30}) into Eq.~(\ref{1.29})
and using Eq.~(\ref{app:3}),
we find the
efficiency of one-photon Fock state generation as
\begin{multline}
  \label{2.5}
  \eta(t) =
  \frac{R_k^2}{4}\frac{\gamma_{k\mathrm{rad}}}{\Gamma}
  \bigg[
    \int_0^t \D t'\, g(t') C_2(t')
	\\ 	
	\times\!
    	\int_0^{t'} \D t''\, g(t'') C_2^*(t'')
        e^{i(\Delta_k + i \Gamma_k/2)(t'-t'')}
  	+
	\mbox{C.c.} 
	\bigg],
\end{multline}
where $\gamma_{k\mathrm{rad}}$ describes the wanted radiative
losses due to transmission of the radiation through the fractionally
transparent mirror.
As we can see,
the efficiency is proportional to the ratio
$\gamma_{k\mathrm{rad}}/\Gamma_k$, where the damping parameter
$\Gamma_k$ can be regarded as being the sum
$\Gamma_k = \gamma_{k\mathrm{rad}} + \gamma_{k}^\prime$, where $\gamma_{k}^\prime$
describes the unwanted losses due to absorption.
For the numerical calculations, we will assume
\mbox{$\gamma_{k\mathrm{rad}}/\Gamma_k$ $\!=0.9$}, which corresponds
to $10\%$ unwanted losses.
Apart from the unavoidably existing unwanted losses,
the one-photon Fock state efficiency depends on the
shape
$g(t)$ and the Rabi frequency $R_k$ of the emitter--cavity
interaction
and the
shape
and the intensity of the external pump field.
Finally, inserting
Eq.~(\ref{1.31}) together with Eq.~(\ref{1.30}) into Eq.~(\ref{1.35}) and
using
Eq.~(\ref{app:3}),
we
find the spatiotemporal shape of the excited outgoing wave packet
outside the cavity as
(in the following, for the sake of simplicity, we assume \mbox{$\omega_{31}$ $\!=0$})
\begin{multline}
  \label{2.1}
  \phi_1(z,t)=
  \frac{ R _k}{2}
 \sqrt{
  {\displaystyle
        \frac{\hbar\omega_k \gamma_{k\mathrm{rad}}}
        {2\varepsilon_0 c\mathcal{A} \eta(t)}
      }}
\\[.5ex] \times\!
     \int_0^{t-z/c}
     \!\D  t'\,
     C_2^*(t')g(t')
     e^{-i(\Delta_k + i \Gamma_k/2) t'}
     e^{i 
     \tilde{\omega}_k^*(t-z/c)
     }
 .
  \end{multline}

To present
explicit results, we restrict
ourselves to
times
much larger than the
time the excited wave packet needs to almost completely exit the
cavity.
To find the spatiotemporal shape of the excited outgoing wave
packet 
$\phi_1(z,T)$ at time \mbox{$T$ $\!\gg$ $\!\Gamma_k^{-1}$}
for
given pump intensity and emitter--cavity interaction shapes
$\Omega_p (t)$ and $g(t)$, respectively,
we first solve
Eq.~(\ref{1.22}) together with Eqs.~(\ref{1.24}), (\ref{1.23}) and
the initial condition \mbox{$C_2(0)=0$}. Equation~(\ref{1.22}) is a
Volterra-type integro-differential equation of the second kind, which
can be solved using a standard fourth-order Runge-Kutta algorithm.
Then, inserting the solution $C_2(t)$ into
Eqs.~(\ref{2.5}) and (\ref{2.1}) we can find, respectively, the
efficiency of one-photon Fock state emission and the spatiotemporal
shape of the wave packet leaving the cavity.

\begin{figure}[t!]
        \includegraphics[width=0.48\textwidth]{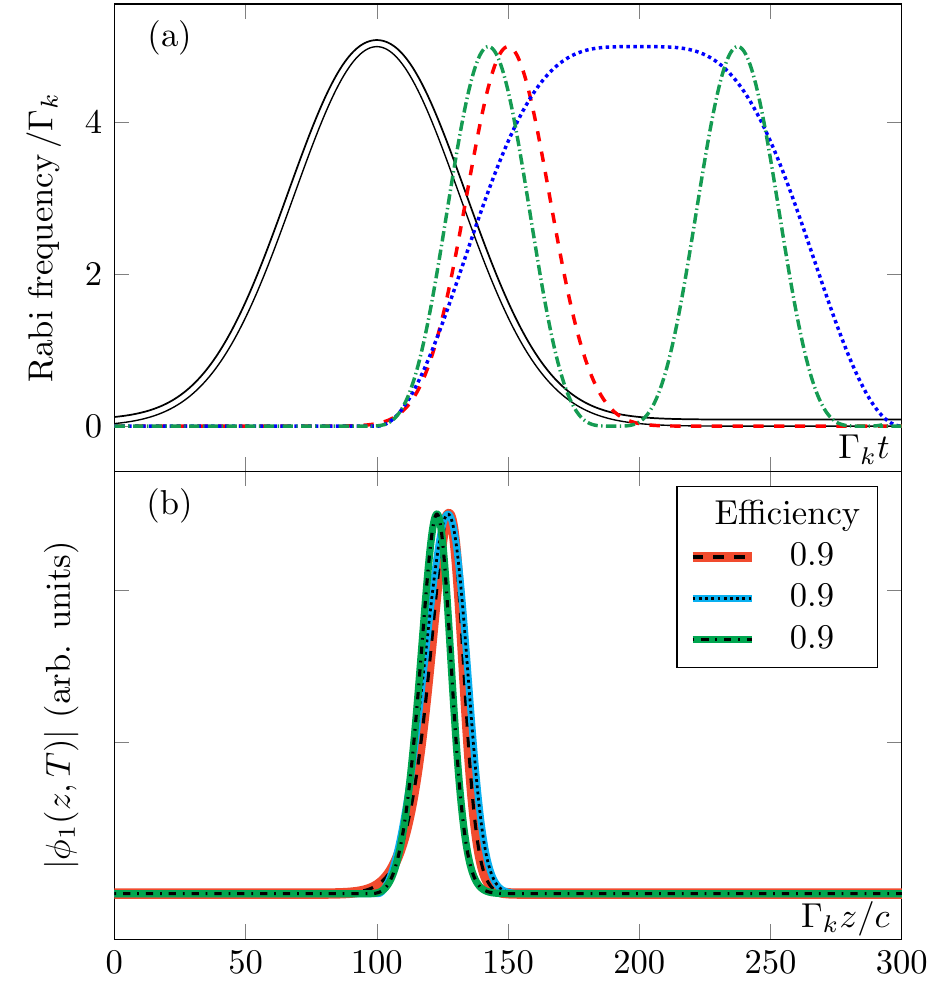}
	\caption{
          (a)
           Various pulse
           shapes
           of the
           pump
           $\Omega_{p}(t)$ (dotted, dashed and
           dash-dotted lines) with the maximum value
           $\Omega_{p\mathrm{max}} = 5\Gamma_k $ and the
           shape
           of the emitter--cavity
           interaction $g(t)$ (double line, shown as normalized to $\Omega_{p\mathrm{max}}$).
           (b) Corresponding spatiotemporal shapes of the 
           one-photon Fock state
           outgoing wave packet
           ($R_k = 5\Gamma_k$,
           $\Delta_k = \Delta_p =  \Gamma_k$, $T = 350 \Gamma_k$).
}
	\label{fig2}
\end{figure}

In accordance
with the vSTIRAP state-mapping proposal~\cite{parkins:3095, law:1055},
in an ideal system
the
maximum one-photon Fock state
efficiency can be achieved when the dynamics of the system follows the
"dark state", which is formed as a superposition of the states
$|\!\left\lbrace0\right\rbrace\!\rangle|1\rangle$ and
$\hat{f}^{\dagger}(z, \omega)|\{0\}\rangle|3\rangle$. This is the case
if the process is adiabatic and the emitter--cavity interaction is
well-established before the driving pulse is
applied.
The case of the vSTIRAP 
is illustrated in Fig.~\ref{fig2}, where a Gaussian-type
emitter--cavity interaction shape and three driving pulse shapes are
considered.   
Comparing the values of $\eta(T)$, 
we see that
\mbox{$\eta(T)$ $\!\approx 0.9$} for all three cases of the driving
pulse shapes, which shows that
the efficiencies of one-photon Fock state generation is
only limited by the unwanted losses in the system.
Moreover, Fig.~\ref{fig2} reveals that the excited outgoing wave packet carrying
a single photon is generated in the time interval immediately as the
driving pulse is applied. Thus,
since the adiabatic evolution required by the vSTIRAP is much longer than
the cavity mode decay time,
the generated single-photon wave packet
escapes from the cavity once it is generated, and, hence,
the shape of the wave packet
$|\phi_1(z,T)|$ is effectively independent of the driving pulse
shape.

%

\subsection{
Effects of emitter--cavity and emitter--pump interaction strengths and shapes
}
\label{sec5.3}
In general,
it is desirable to have
the possibility to control and manipulate
the shape
of the outgoing wave packet
and, at the same time, to
excite
it
in a one-photon Fock state with high efficiency.
Let us therefore study the dependencies of the efficiency and the wave packet shape
on the emitter--cavity interaction and the emitter--pump interction in more detail.
For the sake of transparency
we first concentrate on the case, when
the emitter--cavity interaction is
constant, i.e., it does not vary with time.

\begin{figure}[t!]
	\includegraphics[width=0.48\textwidth]{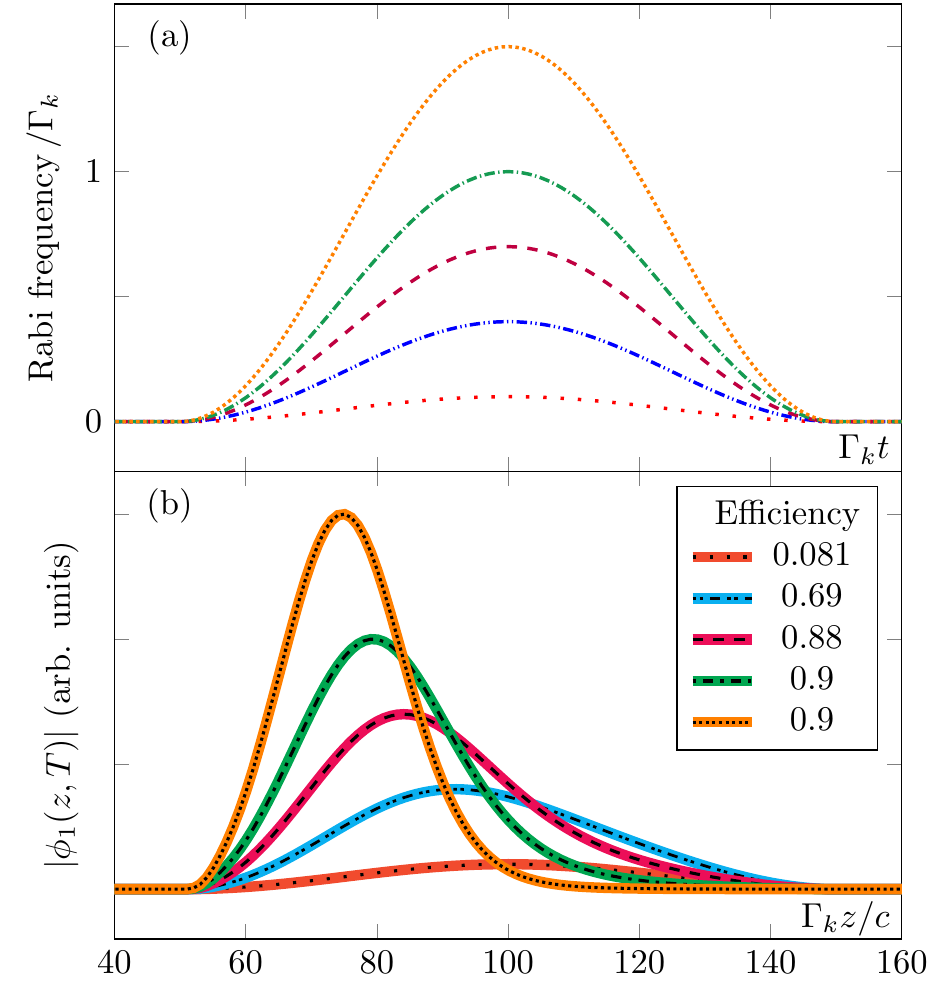}
	\caption{
          (a)
          Various pulse shapes of the pump $\Omega_{p}(t)$
          with the maximum value
           $\Omega_{p\mathrm{max}} =
          0.1
\Gamma_k
          $ (dotted red line),  $\Omega_{p\mathrm{max}} = 0.4
\Gamma_k
          $
          (dashed-dotted blue line),
          $\Omega_{p\mathrm{max}} = 0.7
\Gamma_k
          $ (dashed purple line),
          $\Omega_{p\mathrm{max}} = 1
\Gamma_k
          $ (dashed-dotted green line),
          $\Omega_{p\mathrm{max}} = 1.5
\Gamma_k
          $ (dotted orange line).
          (b) Corresponding spatiotemporal shapes of the
          one-photon Fock state
          outgoing wave packet
          ($R_k = 2\Gamma_k$, $\Delta_k = \Delta_p = 0$, $T = 200 \Gamma_k$).
}
	\label{fig3}
\end{figure}

In Fig.~\ref{fig3}
the spatiotemporal shape of the
single-photon
outgoing wave packet
(solid lines)
is shown
for various intensities of
a symmetric single-peak driving pulse.
It is seen that
when the
value of the
maximum of the driving pulse
is low,
$\Omega_{p\mathrm{max}} = 0.1
\Gamma_k$,
the shape of the outgoing wave packet
(red, solid line in Fig.~\ref{fig3}(b)) perfectly coincides with the
shape
of the driving pulse
(red, dotted line in Fig.~\ref{fig3}(a)).
However, the efficiency of one-photon Fock state generation is quite
low,
$\eta(T)=0.081$.
It is further seen that the efficiency
increases with $\Omega_{p\mathrm{max}}$.
On the other hand, the
figure also reveals that with increasing
$\Omega_{p\mathrm{max}}$
the driving pulse induces the photon emission earlier,
and as a result, the shape of the outgoing wave packet
does not follow the profile of the driving pulse
but
features an asymmetric shape.

\begin{figure}[t!]
	\includegraphics[width=0.48\textwidth]{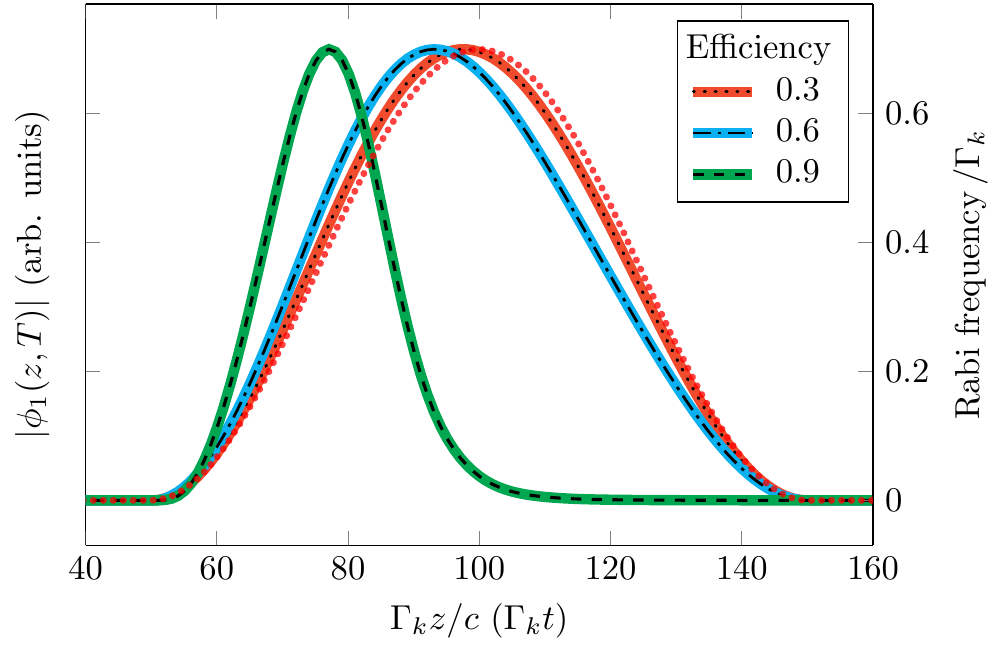}
	\caption{Spatiotemporal shape of the
          one-photon Fock state
          outgoing wave packet
          for
          $R_k = \Gamma_k$ (solid green line), $R_k = 4\Gamma_k$
          (solid blue line),  $R_k = 7\Gamma_k$ (solid red line)
          in the case of a single-peak driving pulse (dotted red line),
          $\Omega_{p\mathrm{max}} = 0.7\Gamma_k$, $\Delta_k = \Delta_p = 0$, $T = 200 \Gamma_k$.}
	\label{fig4}
\end{figure}

Next, let us consider the dependence of the spatiotemporal shape of the
single-photon
outgoing wave packet on the
Rabi frequency of the emitter--cavity interaction $R_k$.
In Fig.~\ref{fig4} the shape
is plotted
for various
values of $R_k$
in the case of a
symmetric
single-peak driving pulse.
For $R_k = \Gamma_k$ the
vSTIRAP case is seen to be realized, resulting in high-efficiency
single-photon Fock state generation and asymmetric spatiotemporal shape
of the associated outgoing wave packet.
For higher values of
$R_k$,
the driving pulse
is not strong enough to realize the
vSTIRAP, and
the efficiency of one-photon Fock state generation is reduced, but
the spatiotemporal shape of the associated wave packet follows
the shape of the driving pulse.

\begin{figure}[t!]
	\includegraphics[width=0.48\textwidth]{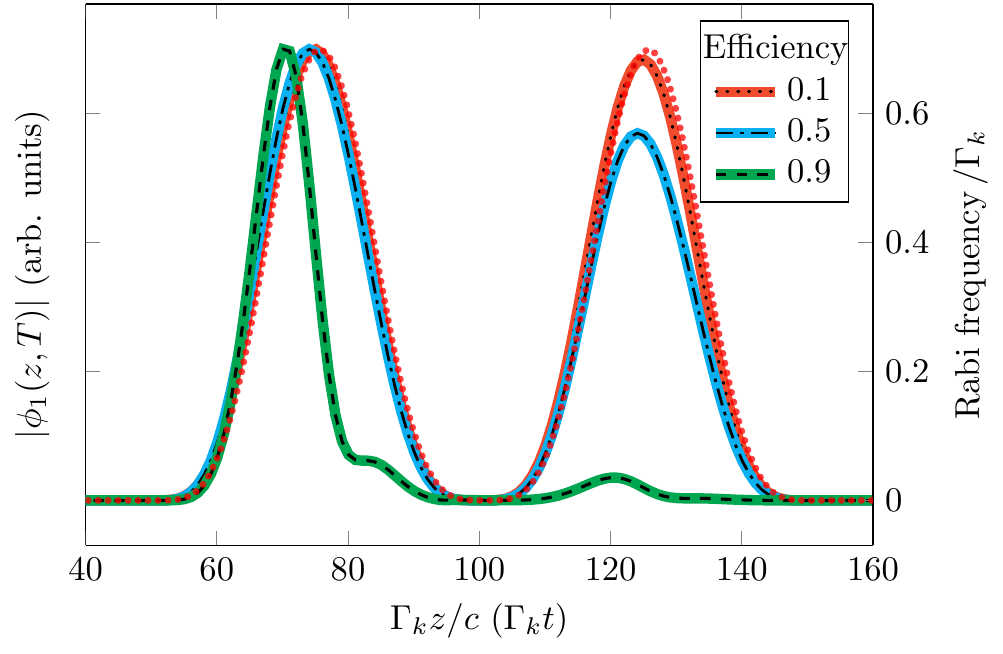}
 	\caption{Spatiotemporal shape of the
          one-photon Fock state
          outgoing wave packet
          for
          $R_k = \Gamma_k$
          (solid green line), $R_k = 4\Gamma_k$ (solid blue line),
          $R_k = 12\Gamma_k$ (solid red line)
          in the case of a double-peak driving pulse (dotted red line),
          $\Omega_{p\mathrm{max}} = 0.7\Gamma_k$, $\Delta_k = \Delta_p = 0$, $T = 200 \Gamma_k$.}
	\label{fig5}
\end{figure}

An analogous situation can be observed for a double-peak driving
pulse as illustrated in Fig.~\ref{fig5}.
For emitter--cavity
Rabi frequencies
of the same order of magnitude as the driving pulse maximum, the
vSTIRAP is
again seen to be
realized and
the efficiency of one-photon Fock state generation is
close to the maximum value of $0.9$,
but the shape of the associated wave packet
becomes asymmetric.
As expected, for
higher rates of emitter--cavity Rabi frequency
it
follows the driving pulse
shape
but the
one-photon Fock state generation efficiency is
rather
low.

\section{
Shape-controlled single-photon wave packet emission
}
\label{sec7}

As we have seen, the
shape
of the
single-photon
outgoing
wave packet
can be changed by changing
the emitter--cavity interaction
and/or
the
driving pulse.
In this context, a question
of particular interest is
the
generation of single-photon wave packets
of controlled shapes.
To give an answer, let us
assume that the desired spatiotemporal shape of the
single-photon
outgoing
wave packet is given by a function $\phi(z,T)$ at
some
time $T$, where
the condition
\mbox{$T$ $\!\gg$ $\!\Gamma_k^{-1}$}
again ensures
that the wave packet
has
almost completely
left
the cavity.

From
Eq.~(\ref{2.1})
[$\phi_1(z,t)$ $\rightarrow$ $\phi(z,T)$] we find that
\begin{align}
  \label{7.1}
  C_2(t) &=
  \frac{2}{R_kg(t)} \sqrt{\frac{2\varepsilon_0c
      \mathcal{A}\eta(T)}{\hbar \omega_k \gamma_{k\mathrm{rad}}}}
  e^{-i \omega_{23} t}
  \nonumber\\[1ex]&
  \times
  \left\lbrace
    \frac{\D \phi[c(T-t),T]}{\D t}
    -i 
    \tilde{\omega}_k^*
    \phi[c(T-t),T]
  \right\rbrace
.
\end{align}
Further, inserting Eq.~(\ref{1.24}) into Eq.~(\ref{1.22}) we see that
\begin{align}
  \label{7.3}
  D(t) = f(t) + \int_0^t \D t' f(t)f(t')C_2(t'),
\end{align}
where $D(t)$ is defined by
\begin{multline}
  \label{7.5}
  D(t) \equiv \dot{C}_2(t)
\\
  + \frac{R_k^2}{4} \int _0^t \D t' C_2(t')
  g(t)g(t')
  e^{-i(\Delta_k - i \Gamma_k/2)(t-t')},
\end{multline}
and $f(t)$ is related to the
shape
of the pump pulse $\Omega_p(t)$ according to
\begin{align}
  \label{7.7}
    f(t) = \frac{i}{2} \Omega_p(t) e^{-i\Delta_pt}.
\end{align}
Differentiation of Eq.~(\ref{7.3}) with respect to $t$ then yields the
following differential equation for $f(t)$:
\begin{align}
  \label{7.9}
  D(t)\dot{f}(t) + C_2(t) f^3(t)-\dot{D}(t)f(t)=0 .
\end{align}

Hence,
the following scheme of shape controlling may be established.
For a chosen (i.e., desired) spatiotemporal shape $\phi(z,T)$
of the single-photon outgoing wave packet, the probability
amplitude $C_2(t)$
is
calculated from Eq.~(\ref{7.1}).
The result is then, together with Eq.~(\ref{7.5}), inserted
into the
differential
equation~(\ref{7.9}).
Its solution eventually
yields the sought
shape
of the driving pulse.

\begin{figure}[t!]
	\includegraphics[width=0.48\textwidth]{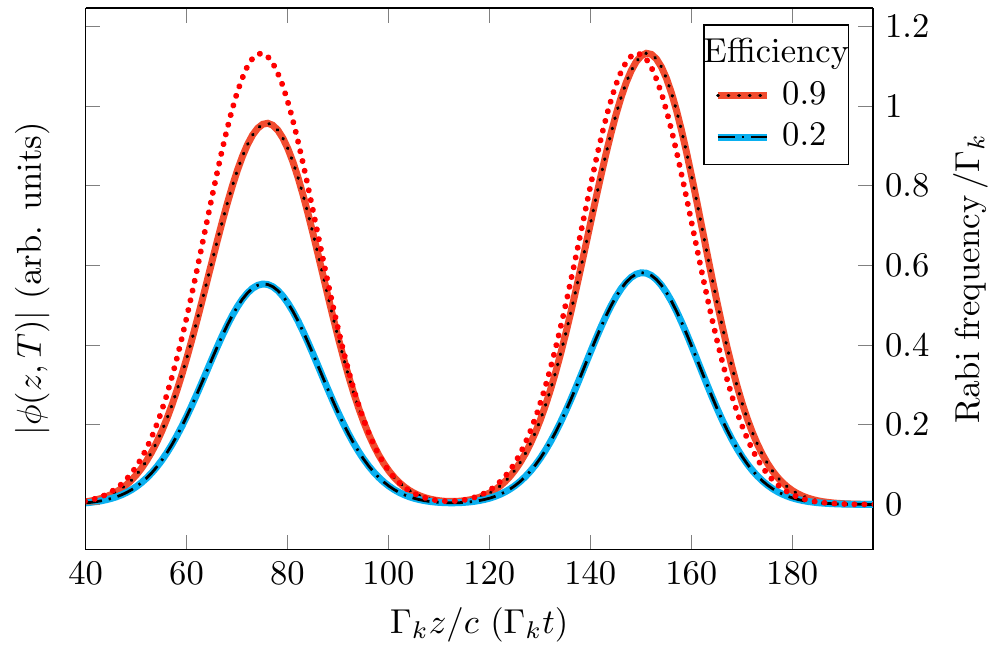}
 	\caption{
          Shapes
          of the driving pulse (solid lines) required to
          generate
          a double-peak wave packet carrying a one-photon Fock state
          (dotted red line) for a high generation efficiency $\eta(T)=0.9$
          (solid red line) and a low generation efficiency $\eta(T)=0.2$ (solid blue line),
          $R_k = 8\Gamma_k$, \mbox{$\Delta_k = \Delta_p = 0$}, \mbox{$T = 200 \Gamma_k$}.}
	\label{fig7}
\end{figure}

To illustrate the
scheme,
we
apply it first to
the generation of
single-photon double-peak wave packets
in the case, when the emitter--cavity interaction is constant,
as shown in
Fig.~\ref{fig7} for two one-photon Fock state generation efficiencies.
In the case of low efficiency the shape of the single-photon
outgoing wave packet is seen to closely follow
the shape of the driving pulse. It is also seen that
when an efficiency close to
unity is required, then the second peak of the driving pulse must be
higher than that of the first one.
This can be explained by the fact that during the generation of
a single-photon wave packet, a part of the wave packet (which
corresponds to the first peak)
already escapes
the cavity. Therefore, to generate the second required peak the
pump intensity should be higher.

\begin{figure}[t!]
	\includegraphics[width=0.48\textwidth]{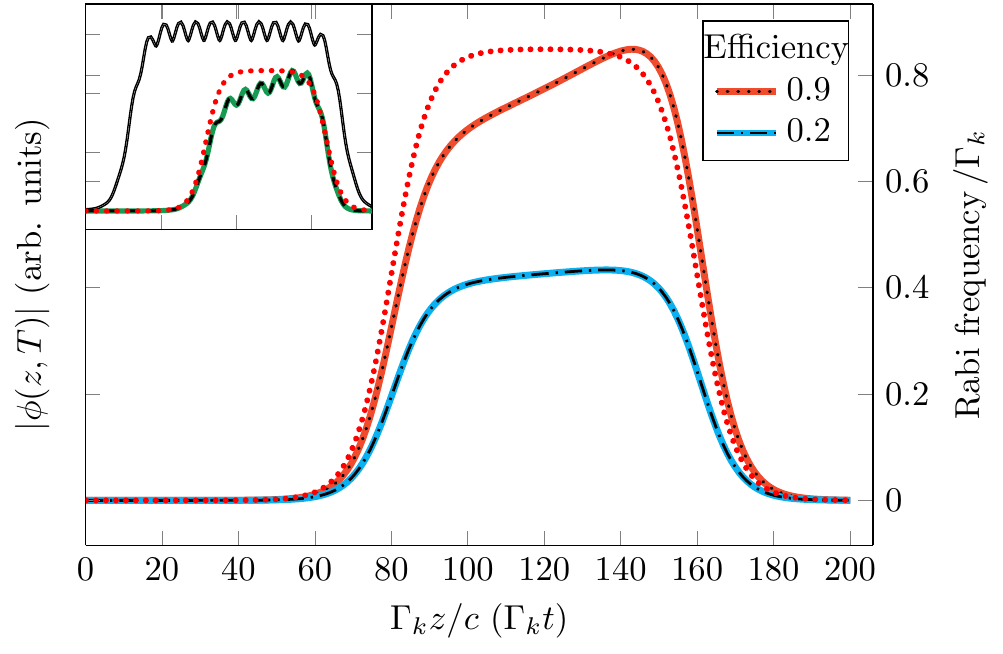}
 	\caption{
          Shapes
          of the driving pulse (solid lines) required to
          generate
          a plateau-like wave packet carrying a one-photon Fock state
          (dotted red line) for a high generation efficiency $\eta(T)=0.9$
          (solid red line) and a low generation efficieny
          $\eta(T)=0.2$ (solid blue line), $R_k = 8\Gamma_k$, \mbox{$\Delta_k =
          \Delta_p = 0$, $T = 200 \Gamma_k$}. The inset shows the required
          shape
          of the driving pulse (solid green line) in the case of a time-dependent
          shape of the emitter--cavity interaction
          (solid black line).}
	\label{fig8}
\end{figure}

A similar behavior is observed in the case when a plateau-like
spatiotemporal shape of the single-photon wave packet is desired.
as it can be seen from Fig.~\ref{fig8}.
Namely, in the case of constant emitter--cavity interaction,  
for low efficiencies the required profile of the driving
pulse is similar to the spatiotemporal shape of the outgoing
radiation. For generation efficiencies close to unity, the
shape
of the driving
pulse needs to be asymmetric.
For comparison, a case of time-dependent emitter--cavity interaction
is shown in the inset of Fig.~\ref{fig8}.
In particular, the interaction shape is chosen to simulate 
time-dependent variation of the location of the emitter in the cavity, 
 as
small oscillations of the interaction due to the variation of emitter
position around the anti-nodes of the cavity mode.
It is seen that to
obtain the desired shape of the single-photon outgoing wave packet,
the profile of the driving pulse should follow the variation of the
emitter--cavity interaction.
%
%
%

\section{Summary and concluding remarks}
\label{sec9}

Based on macroscopic QED in dispersing and absorbing media, we
have presented an exact description of the resonant interaction of a
three-level $\Lambda$-type emitter in a high-$Q$ cavity with the
cavity-assisted electromagnetic field.
In particular, we have studied the case when one transition
of the emitter resonantly interacts with the cavity field and the
other one is subjected to an external classical pump field.
We have applied the theory to determine
both the efficiency of one-photon Fock state generation
and the spatiotemporal shape of the outgoing wave packet.
In this context, we have shown that,
even for high one-photon Fock state generation efficiencies,
wave packets of arbitrary desired shape can be generated by
appropriately adjusting the pump.

Under the
assumption that the Hilbert space of the total
system is spanned by
a single-quantum excitation,
the quantum state of the
excited outgoing nonmonochromatic mode is always a mixture of
a one-photon Fock state and
the vacuum state. Due to the unavoidable existence of unwanted losses,
such as absorption or scattering, the efficiency of one-photon Fock
state generation
is always limited by the ratio of
the cavity radiative decay rate to the total decay rate, which
includes both wanted
and unwanted losses.

The
spatiotemporal shape of the excited outgoing wave packet
and
the efficiency of one-photon Fock state generation
depend on
the interplay of various parameters
of the emitter--cavity and emitter--pump interactions, such as
their strengths
and
shapes.
In particular, in the case
of the
vSTIRAP
the efficiency of one-photon Fock state generation may
be close to the upper limit
determined
by the unavoidable
unwanted losses in the system.
It
requires that the process is adiabatic
and the emitter--cavity interaction is well-established before the
driving pulse is applied---so that the system dynamics follows
the dark state.
However,
in this regime,
control of the spatiotemporal shape
of the excited outgoing wave packet is not possible, as the shape is
nearly
independent of the driving pulse shape.

In the case when the emitter-cavity interaction can be regarded as
being time independent on the timescales of the cavity decay time and
the driving pulse length, two interaction regimes may be identified.
In the regime
when the emitter--cavity interaction
Rabi frequency
is
sufficiently high compared to
the (maximum) emitter--pump interaction
Rabi frequency,
the spatiotemporal
shape
of the excited outgoing wave packet follows the
shape
of the driving pulse.
However, the one-photon Fock state generation efficiency
is very low. This interaction regime
has been realized by means of the interaction of a high-finesse cavity with
both neutral atoms~\cite{kuhn:067901} and trapped ions~\cite{keller:1075},
where
single-photon wave packet
shapes
that follow the driving field
shapes
have been reported.
In the regime
when the
emitter--pump interaction
Rabi frequency
is of the same order of magnitude as the emitter--cavity
Rabi frequency,
the one-photon Fock state generation efficiency is close to the upper limit.
However, the spatiotemporal shape of the associated outgoing wave packet
has a single-peak profile
and is
independent on the driving pulse profile.

Desired shapes of outgoing wave packet 
with high efficiency of one-photon Fock state
can be generated
by appropriately adjusting 
the emitter--pump interaction
for given emitter--cavity interaction.
Since the
spatiotemporal shape of the wave packet is robust against
fluctuations of 
the pump pulse,
the results found imply the feasibility
to produce identical wave packets of desired shape
carrying a one-photon Fock state each.
As an
illustration of the
scheme,
we have considered the generation of
a single-photon double-peak wave packet.
In fact,
the one-photon Fock state is equally distributed
among
two time-(space-)separated
single-peak wave packets, which is a realization of a
time-bin entanglement that can be used to perform quantum key
distribution~\cite{tittel:4737}.
Importantly,
the scheme also allows an
optimization of the pump pulse profile
to match
with specific characteristics of
the time dependence of the emitter--cavity interaction,
such as injecting
the emitter into the cavity
or fluctuating emitter motion
around the center of the interaction region.

\begin{acknowledgments}
  M.~K. acknowledges valuable discussions with J.~Wiersig.
\end{acknowledgments}

\appendix

\section{Green tensor }
\label{app}
   The Green tensor has the following property:
   \begin{equation}
     \label{app:1}
     \frac {\omega ^2} {c^2}\!
     \int  \D x\,
      \varepsilon''(x, \omega)
     G(z_1, x, \omega)G^*(z_2, x, \omega)
    =
     \mathrm{Im}\,G
     (z_1, z_2,\omega) .
   \end{equation}
The Green tensors $G (z_A, z_A, \omega )$
and $G (0^+, z_A, \omega )$
read (see, Ref.~\cite{khanbekyan:013822})
\begin{equation}
  \label{app:2}
   G (z_A, z_A, \omega )
    = -
   \sum _k
   \frac{c^2}{ \tilde{\omega}_kl}
   \frac{1} {\omega-\tilde{\omega}_k}
   \sin ^2( \omega_k  z_A/c)
\end{equation}
and
\begin{equation}
  \label{app:3}
   G (0^+, z_A, \omega )
    = -
   \sum _k
   \frac{t_{13}c^2}{2 \tilde{\omega}_k l}
   \frac{i} {\omega-\tilde{\omega}_k}
   e^{i\omega_k l/c}
   \sin ( \omega_k  z_A/c),
\end{equation}
respectively,
where $t_{13}$ denotes the transmission coefficient from inside the
cavity to outside.


\section{Quantum State of the Outgoing field}
\label{app2}

To calculate the quantum state of the outgoing field, following
Ref.~\cite{khanbekyan:013822}, we start from
the one-dimensional version of Eq.~(\ref{1.11}) and decompose the
electric field 
into incoming and outgoing fields. Then, we may represent the outgoing
field at the point $z$ $\!=$ $\!0$ outside the cavity as
\begin{multline}
      \label{app2.1}
     \uh{ E}_{\mathrm{out}}(z,\omega)
       \bigr|_{z=0^+}
      = i \sqrt{\frac {\hbar }{\varepsilon _0\pi \mathcal{A}}}\,
      \frac{\omega^2}{c^2}
\\[.5ex]
\times
      \int \D z'\sqrt{\varepsilon''(z',\omega)}\,
      G_{\rm out}(0^+,z',\omega)
      \hat{f}(z',\omega).
\end{multline}
It is useful to introduce
the bosonic operators
\begin{equation}
      \label{app2.3}
      \hat{ b}_{\mathrm{out}}  (\omega)
      = 2\,\sqrt{\frac{\varepsilon_0 c \pi\mathcal{A}}{\hbar\omega}}
     \,\left.\uh{ E}_{\mathrm{out}}(z, \omega)
     \right|_{z=0^+}
,
\end{equation}
where
\begin{equation}
  \label{app2.5}
  \bigl[\hat{b}_{\mathrm{out}}(\omega),
       \hat{b}_{\mathrm{out}}^\dagger(\omega')\bigr]
       =
\delta (\omega - \omega ').
\end{equation}
To calculate the quantum state of the outgoing field,
we start from the multimode characteristic functional
\begin{multline}
\label{app2.7}
C_{{\rm out}}[\beta(\omega),t]
\\[.5ex]
         =
         \left
          \langle
           \psi(t)
         \right|
         \exp\!\left[
         \int_0^{\infty}\!
         \D\omega\,
         \beta(\omega) \hat{b}^{\dagger}_{\mathrm {out}}(\omega)
         - \mathrm{H.c.}\right]\!
       \left|
       \psi(t)
        \right
        \rangle
.
\end{multline}
To evaluate $C_{{\rm out}}[\beta(\omega),t]$ for the
state $|\psi(t)\rangle$ as given by Eq.~(\ref{1.17}), we
first note that from Eq.~(\ref{1.17}) together with the relation
$\hat{f}(z, \omega)|\lbrace 0\rbrace\rangle$
$\!=$ $\!0$ it follows that
\begin{equation}
  \label{app2.9}
   \hat{f}(z, \omega) |\psi(t)\rangle =
    C_3(z, \omega, t) 
    e^{-i(\omega+\omega_{31}) t}\
   |\lbrace 0\rbrace\rangle
|3\rangle
   .
\end{equation}
Hence, on recalling Eqs.~(\ref{app2.1}) and (\ref{app2.3}),
it can be seen that
\begin{equation}
  \label{app2.11}
  \hat{ b}_{\mathrm{out}}  (\omega)
   |\psi(t)\rangle =
    F^*(\omega, t)
   \
   |\lbrace 0\rbrace\rangle
   |3\rangle
   ,
\end{equation}
where
\begin{multline}
  \label{app2.13}
    F(\omega, t)=
    -2i
   \sqrt{\frac{c}{\omega}}
    \frac{\omega^2}{c^2}
     e^{i(\omega+\omega_{31}) t}
\\[.5ex]
\times
      \int \D z\,
       \sqrt{\varepsilon''(z, \omega)}\,
     G^*_{\rm out}(0^+,z,\omega)
     C_3^*(z,\omega, t)
.
\end{multline}
To represent $C_{\mathrm{out}}[\beta(\omega),t]$ in a more transparent
form, we introduce a time-dependent
unitary transformation according to
\begin{align}
  \label{app2.15}
&
        \beta(\omega)
         =
      \sum_i F_i^\ast
(\omega,t)
\beta_i(t),
\\[.5ex] \label{app2.17}
&
\beta_i(t)
   =
   \int_0^{\infty}\!
    \D\omega
    F_i(\omega,t)
          \beta (\omega )
.
\end{align}
Here, the normalized
complex functions \mbox{$F_i(\omega,t)$} represent nonmonochromatic
modes of appropriately chosen frequency distributions where associated
nonmonochromatic mode operators read
 \begin{equation}
  \label{app2.19}
 \hat{b}_{\mathrm{out}\,i}(t)
      = \int_0^{\infty}\!
      \D\omega\,
    F_i(\omega,t)\hat{b}_{\mathrm {out}} (\omega).
\end{equation}
Note that
\begin{equation}
  \label{app2.21}
\hat{b}_{\mathrm {out}} (\omega)
=
  \sum_i
  F_i^\ast(\omega,t)
  \hat{b}_{\mathrm{out}\,i}(t)
.
\end{equation}
Accordingly, applying the Baker-Campberll-Hausdorf formula, using the
commutation relation Eq.~(\ref{app2.5}) together with
Eq.~(\ref{app2.11}) we may rewrite
$C_{\mathrm{out}}[\beta(\omega),t]$, Eq.~(\ref{app2.7}) as
\begin{multline}
  \label{app2.23}
   C_{\mathrm{out}}[\beta_i(t),t]=
    \exp\left[ -{\textstyle\frac{1}{2}}
    \sum_i
      |\beta_i(t)|^2
    \right]
\\[.5ex]
    \times
    \left[
      1-\left|\sum_i \beta_i(t)
     \int_0^{\infty}\!
     \D\omega\,
     F_i(\omega,t)F(\omega, t)\right|^2
    \right].
\end{multline}
We now choose
\begin{equation}
  \label{app2.25}
    F_1(\omega,t)
    = \frac{F(\omega , t)}{\sqrt{\eta(t)}}\,,
\end{equation}
with $\eta(t)$ being given by Eq.~(\ref{1.29}).
In this way, from Eq.~(\ref{app2.23}) we obtain
$C_{\mathrm{out}}[\beta_i(t),t]$ in a 'diagonal' form
with respect to the nonmonochromatic modes:
\begin{equation}
  \label{app2.27}
C_\mathrm{out}[\beta_i(t),t]
= C_1[\beta_1(t),t]\prod_{i\neq1} C_i[\beta_i(t),t],
\end{equation}
where
\begin{equation}
  \label{app2.29}
 C_1(\beta,t) = e^{-|\beta|^2/2}
\left[1-\eta(t)|\beta|^2\right]
\end{equation}
and
\begin{equation}
  \label{5.21-3}
 C_i(\beta,t) = e^{-|\beta|^2/2} \quad (i\neq 1).
\end{equation}
The Fourier transform of $C_\mathrm{out}[\beta_i(t),t]$
with respect to the $\beta_i(t)$ then yields the (multi-mode)
Wigner function $W_{\mathrm{out}}(\alpha_i,t)$, Eq.~(\ref{1.25}).

Substitution of the formal solution of Eq.~(\ref{1.18}) for
$C_3(z, \omega, t)$
[with the initial condition 
\mbox{$C_3(z, \omega, 0)$ $\!=$ $\!0$}] into Eq.~(\ref{app2.13}) and
the use of Eq.~(\ref{app:1}) yields Eq.~(\ref{1.30}) 
(for details, see Ref.~\cite{khanbekyan:013822}).


\begin{thebibliography}{21}
\expandafter\ifx\csname natexlab\endcsname\relax\def\natexlab#1{#1}\fi
\expandafter\ifx\csname bibnamefont\endcsname\relax
  \def\bibnamefont#1{#1}\fi
\expandafter\ifx\csname bibfnamefont\endcsname\relax
  \def\bibfnamefont#1{#1}\fi
\expandafter\ifx\csname citenamefont\endcsname\relax
  \def\citenamefont#1{#1}\fi
\expandafter\ifx\csname url\endcsname\relax
  \def\url#1{\texttt{#1}}\fi
\expandafter\ifx\csname urlprefix\endcsname\relax\def\urlprefix{URL }\fi
\providecommand{\bibinfo}[2]{#2}
\providecommand{\eprint}[2][]{\url{#2}}

\bibitem[{\citenamefont{Reiserer and Rempe}(2015)}]{reiserer:1379}
\bibinfo{author}{\bibfnamefont{A.}~\bibnamefont{Reiserer}} \bibnamefont{and}
  \bibinfo{author}{\bibfnamefont{G.}~\bibnamefont{Rempe}},
  \bibinfo{journal}{Rev. Mod. Phys.} \textbf{\bibinfo{volume}{87}},
  \bibinfo{pages}{1379} (\bibinfo{year}{2015}).

\bibitem[{\citenamefont{Knill et~al.}(2001)\citenamefont{Knill, Laflamme, and
  Milburn}}]{knill:46}
\bibinfo{author}{\bibfnamefont{E.}~\bibnamefont{Knill}},
  \bibinfo{author}{\bibfnamefont{R.}~\bibnamefont{Laflamme}}, \bibnamefont{and}
  \bibinfo{author}{\bibfnamefont{G.~J.} \bibnamefont{Milburn}},
  \bibinfo{journal}{Nature} \textbf{\bibinfo{volume}{409}}, \bibinfo{pages}{46}
  (\bibinfo{year}{2001}).

\bibitem[{\citenamefont{Cirac et~al.}(1997)\citenamefont{Cirac, Zoller, Kimble,
  and Mabuchi}}]{cirac:3221}
\bibinfo{author}{\bibfnamefont{J.~I.} \bibnamefont{Cirac}},
  \bibinfo{author}{\bibfnamefont{P.}~\bibnamefont{Zoller}},
  \bibinfo{author}{\bibfnamefont{H.~J.} \bibnamefont{Kimble}},
  \bibnamefont{and} \bibinfo{author}{\bibfnamefont{H.}~\bibnamefont{Mabuchi}},
  \bibinfo{journal}{Phys.\ Rev.\ Lett.} \textbf{\bibinfo{volume}{78}},
  \bibinfo{pages}{3221} (\bibinfo{year}{1997}).

\bibitem[{\citenamefont{Marcikic et~al.}(2002)\citenamefont{Marcikic,
  de~Riedmatten, Tittel, Scarani, Zbinden, and Gisin}}]{marcikic:062308}
\bibinfo{author}{\bibfnamefont{I.}~\bibnamefont{Marcikic}},
  \bibinfo{author}{\bibfnamefont{H.}~\bibnamefont{de~Riedmatten}},
  \bibinfo{author}{\bibfnamefont{W.}~\bibnamefont{Tittel}},
  \bibinfo{author}{\bibfnamefont{V.}~\bibnamefont{Scarani}},
  \bibinfo{author}{\bibfnamefont{H.}~\bibnamefont{Zbinden}}, \bibnamefont{and}
  \bibinfo{author}{\bibfnamefont{N.}~\bibnamefont{Gisin}},
  \bibinfo{journal}{Phys. Rev. A} \textbf{\bibinfo{volume}{66}},
  \bibinfo{pages}{062308} (\bibinfo{year}{2002}).

\bibitem[{\citenamefont{Law and Kimble}(1997)}]{law:2067}
\bibinfo{author}{\bibfnamefont{C.~K.} \bibnamefont{Law}} \bibnamefont{and}
  \bibinfo{author}{\bibfnamefont{H.~J.} \bibnamefont{Kimble}},
  \bibinfo{journal}{J.\ Mod.\ Opt.} \textbf{\bibinfo{volume}{44}},
  \bibinfo{pages}{2067} (\bibinfo{year}{1997}).

\bibitem[{\citenamefont{Bergmann et~al.}(1998)\citenamefont{Bergmann, Theuer,
  and Shore}}]{bergmann:1003}
\bibinfo{author}{\bibfnamefont{K.}~\bibnamefont{Bergmann}},
  \bibinfo{author}{\bibfnamefont{H.}~\bibnamefont{Theuer}}, \bibnamefont{and}
  \bibinfo{author}{\bibfnamefont{B.~W.} \bibnamefont{Shore}},
  \bibinfo{journal}{Rev. Mod. Phys.} \textbf{\bibinfo{volume}{70}},
  \bibinfo{pages}{1003} (\bibinfo{year}{1998}).

\bibitem[{\citenamefont{Specht et~al.}(2009)\citenamefont{Specht, Bochmann,
  Mucke, Weber, Figueroa, Moehring, and Rempe}}]{specht:469}
\bibinfo{author}{\bibfnamefont{H.~P.} \bibnamefont{Specht}},
  \bibinfo{author}{\bibfnamefont{J.}~\bibnamefont{Bochmann}},
  \bibinfo{author}{\bibfnamefont{M.}~\bibnamefont{M\"ucke}},
  \bibinfo{author}{\bibfnamefont{B.}~\bibnamefont{Weber}},
  \bibinfo{author}{\bibfnamefont{E.}~\bibnamefont{Figueroa}},
  \bibinfo{author}{\bibfnamefont{D.~L.} \bibnamefont{Moehring}},
  \bibnamefont{and} \bibinfo{author}{\bibfnamefont{G.}~\bibnamefont{Rempe}},
  \bibinfo{journal}{Nat Photon} \textbf{\bibinfo{volume}{3}},
  \bibinfo{pages}{469} (\bibinfo{year}{2009}).

\bibitem[{\citenamefont{Kiraz et~al.}(2004)\citenamefont{Kiraz, Atat\"ure, and
  Imamo\u{g}lu}}]{kiraz:032305}
\bibinfo{author}{\bibfnamefont{A.}~\bibnamefont{Kiraz}},
  \bibinfo{author}{\bibfnamefont{M.}~\bibnamefont{Atat\"ure}},
  \bibnamefont{and}
  \bibinfo{author}{\bibfnamefont{A.}~\bibnamefont{Imamo\u{g}lu}},
  \bibinfo{journal}{Phys. Rev. A} \textbf{\bibinfo{volume}{69}},
  \bibinfo{pages}{032305} (\bibinfo{year}{2004}).

\bibitem[{\citenamefont{Jaritz and Hohenester}(2011)}]{jaritz:29}
\bibinfo{author}{\bibfnamefont{G.}~\bibnamefont{Jaritz}} \bibnamefont{and}
  \bibinfo{author}{\bibfnamefont{U.}~\bibnamefont{Hohenester}},
  \bibinfo{journal}{Eur. Phys. J. B} \textbf{\bibinfo{volume}{82}},
  \bibinfo{pages}{29} (\bibinfo{year}{2011}).

\bibitem[{\citenamefont{M\"ucke et~al.}(2013)\citenamefont{M\"ucke, Bochmann,
  Hahn, Neuzner, N\"olleke, Reiserer, Rempe, and Ritter}}]{muecke:063805}
\bibinfo{author}{\bibfnamefont{M.}~\bibnamefont{M\"ucke}},
  \bibinfo{author}{\bibfnamefont{J.}~\bibnamefont{Bochmann}},
  \bibinfo{author}{\bibfnamefont{C.}~\bibnamefont{Hahn}},
  \bibinfo{author}{\bibfnamefont{A.}~\bibnamefont{Neuzner}},
  \bibinfo{author}{\bibfnamefont{C.}~\bibnamefont{N\"olleke}},
  \bibinfo{author}{\bibfnamefont{A.}~\bibnamefont{Reiserer}},
  \bibinfo{author}{\bibfnamefont{G.}~\bibnamefont{Rempe}}, \bibnamefont{and}
  \bibinfo{author}{\bibfnamefont{S.}~\bibnamefont{Ritter}},
  \bibinfo{journal}{Phys. Rev. A} \textbf{\bibinfo{volume}{87}},
  \bibinfo{pages}{063805} (\bibinfo{year}{2013}).

\bibitem[{\citenamefont{Kuhn et~al.}(2002)\citenamefont{Kuhn, Hennrich, and
  Rempe}}]{kuhn:067901}
\bibinfo{author}{\bibfnamefont{A.}~\bibnamefont{Kuhn}},
  \bibinfo{author}{\bibfnamefont{M.}~\bibnamefont{Hennrich}}, \bibnamefont{and}
  \bibinfo{author}{\bibfnamefont{G.}~\bibnamefont{Rempe}},
  \bibinfo{journal}{Phys.\ Rev.\ Lett.} \textbf{\bibinfo{volume}{89}},
  \bibinfo{pages}{067901} (\bibinfo{year}{2002}).

\bibitem[{\citenamefont{Keller et~al.}(2004)\citenamefont{Keller, Lange,
  Hayasaka, Lange, and Walther}}]{keller:1075}
\bibinfo{author}{\bibfnamefont{M.}~\bibnamefont{Keller}},
  \bibinfo{author}{\bibfnamefont{B.}~\bibnamefont{Lange}},
  \bibinfo{author}{\bibfnamefont{K.}~\bibnamefont{Hayasaka}},
  \bibinfo{author}{\bibfnamefont{W.}~\bibnamefont{Lange}}, \bibnamefont{and}
  \bibinfo{author}{\bibfnamefont{H.}~\bibnamefont{Walther}},
  \bibinfo{journal}{Nature} \textbf{\bibinfo{volume}{431}},
  \bibinfo{pages}{1075} (\bibinfo{year}{2004}).

\bibitem[{\citenamefont{Vasilev et~al.}(2010)\citenamefont{Vasilev, Ljunggren,
  and Kuhn}}]{vasilev:063024}
\bibinfo{author}{\bibfnamefont{G.~S.} \bibnamefont{Vasilev}},
  \bibinfo{author}{\bibfnamefont{D.}~\bibnamefont{Ljunggren}},
  \bibnamefont{and} \bibinfo{author}{\bibfnamefont{A.}~\bibnamefont{Kuhn}},
  \bibinfo{journal}{New Journal of Physics} \textbf{\bibinfo{volume}{12}},
  \bibinfo{pages}{063024} (\bibinfo{year}{2010}).

\bibitem[{\citenamefont{Nisbet-Jones et~al.}(2011)\citenamefont{Nisbet-Jones,
  Dilley, Ljunggren, and Kuhn}}]{nisbet-jones:103036}
\bibinfo{author}{\bibfnamefont{P.~B.~R.} \bibnamefont{Nisbet-Jones}},
  \bibinfo{author}{\bibfnamefont{J.}~\bibnamefont{Dilley}},
  \bibinfo{author}{\bibfnamefont{D.}~\bibnamefont{Ljunggren}},
  \bibnamefont{and} \bibinfo{author}{\bibfnamefont{A.}~\bibnamefont{Kuhn}},
  \bibinfo{journal}{New J. Phys.} \textbf{\bibinfo{volume}{13}},
  \bibinfo{pages}{103036} (\bibinfo{year}{2011}).

\bibitem[{\citenamefont{Khanbekyan et~al.}(2008)\citenamefont{Khanbekyan,
  Welsch, \surname{Di Fidio}, and Vogel}}]{khanbekyan:013822}
\bibinfo{author}{\bibfnamefont{M.}~\bibnamefont{Khanbekyan}},
  \bibinfo{author}{\bibfnamefont{D.-G.} \bibnamefont{Welsch}},
  \bibinfo{author}{\bibfnamefont{C.}~\bibnamefont{\surname{Di Fidio}}},
  \bibnamefont{and} \bibinfo{author}{\bibfnamefont{W.}~\bibnamefont{Vogel}},
  \bibinfo{journal}{Phys.\ Rev.\ A} \textbf{\bibinfo{volume}{78}},
  \bibinfo{eid}{013822} (\bibinfo{year}{2008}).

\bibitem[{\citenamefont{\surname{Di Fidio}
  et~al.}(2008)\citenamefont{\surname{Di Fidio}, Vogel, Khanbekyan, and
  Welsch}}]{fidio:043822}
\bibinfo{author}{\bibfnamefont{C.}~\bibnamefont{\surname{Di Fidio}}},
  \bibinfo{author}{\bibfnamefont{W.}~\bibnamefont{Vogel}},
  \bibinfo{author}{\bibfnamefont{M.}~\bibnamefont{Khanbekyan}},
  \bibnamefont{and} \bibinfo{author}{\bibfnamefont{D.-G.}
  \bibnamefont{Welsch}}, \bibinfo{journal}{Phys.\ Rev.\ A}
  \textbf{\bibinfo{volume}{77}}, \bibinfo{eid}{043822} (\bibinfo{year}{2008}).

\bibitem[{\citenamefont{Kn\"{o}ll et~al.}(2001)\citenamefont{Kn\"{o}ll, Scheel,
  and Welsch}}]{knoell:1}
\bibinfo{author}{\bibfnamefont{L.}~\bibnamefont{Kn\"{o}ll}},
  \bibinfo{author}{\bibfnamefont{S.}~\bibnamefont{Scheel}}, \bibnamefont{and}
  \bibinfo{author}{\bibfnamefont{D.-G.} \bibnamefont{Welsch}},
  \emph{\bibinfo{title}{Coherence and Statistics of Photons and Atoms}}
  (\bibinfo{publisher}{Wiley}, \bibinfo{address}{New York},
  \bibinfo{year}{2001}), p.~\bibinfo{pages}{1}, \eprint{quant-ph/0003121}.

\bibitem[{\citenamefont{Vogel and Welsch}(2006)}]{vogel}
\bibinfo{author}{\bibfnamefont{W.}~\bibnamefont{Vogel}} \bibnamefont{and}
  \bibinfo{author}{\bibfnamefont{D.-G.} \bibnamefont{Welsch}},
  \emph{\bibinfo{title}{Quantum Optics}} (\bibinfo{publisher}{Wiley-VCH},
  \bibinfo{address}{Weinheim}, \bibinfo{year}{2006}), \bibinfo{edition}{3rd}
  ed.

\bibitem[{\citenamefont{Parkins et~al.}(1993)\citenamefont{Parkins, Marte,
  Zoller, and Kimble}}]{parkins:3095}
\bibinfo{author}{\bibfnamefont{A.~S.} \bibnamefont{Parkins}},
  \bibinfo{author}{\bibfnamefont{P.}~\bibnamefont{Marte}},
  \bibinfo{author}{\bibfnamefont{P.}~\bibnamefont{Zoller}}, \bibnamefont{and}
  \bibinfo{author}{\bibfnamefont{H.~J.} \bibnamefont{Kimble}},
  \bibinfo{journal}{Phys.\ Rev.\ Lett.} \textbf{\bibinfo{volume}{71}},
  \bibinfo{pages}{3095} (\bibinfo{year}{1993}).

\bibitem[{\citenamefont{Law and Eberly}(1996)}]{law:1055}
\bibinfo{author}{\bibfnamefont{C.~K.} \bibnamefont{Law}} \bibnamefont{and}
  \bibinfo{author}{\bibfnamefont{J.~H.} \bibnamefont{Eberly}},
  \bibinfo{journal}{Phys.\ Rev.\ Lett.} \textbf{\bibinfo{volume}{76}},
  \bibinfo{pages}{1055} (\bibinfo{year}{1996}).

\bibitem[{\citenamefont{Tittel et~al.}(2000)\citenamefont{Tittel, Brendel,
  Zbinden, and Gisin}}]{tittel:4737}
\bibinfo{author}{\bibfnamefont{W.}~\bibnamefont{Tittel}},
  \bibinfo{author}{\bibfnamefont{J.}~\bibnamefont{Brendel}},
  \bibinfo{author}{\bibfnamefont{H.}~\bibnamefont{Zbinden}}, \bibnamefont{and}
  \bibinfo{author}{\bibfnamefont{N.}~\bibnamefont{Gisin}},
  \bibinfo{journal}{Phys. Rev. Lett.} \textbf{\bibinfo{volume}{84}},
  \bibinfo{pages}{4737} (\bibinfo{year}{2000}).

\end{thebibliography}

\end{document}